\documentclass{article}
\usepackage{amsmath,amscd}
\usepackage{graphicx}
\usepackage{algorithmicx}
\usepackage{algorithm}
\usepackage{algpseudocode}
\usepackage{amssymb,array}
\usepackage{multirow}
\usepackage{multicol}
\usepackage{spconf}
\usepackage{bm}
\usepackage{xcolor}
\usepackage{caption}
\usepackage{subcaption}
\usepackage{url}

\def\x{\bm{x}}
\def\y{\bm{y}}
\def\w{\bm{w}}
\def\A{\bm{A}}
\def\z{\bm{z}}
\def\r{\bm{r}}
\def\W{\bm{W}}
\def\F{\bm{F}}
\def\S{\bm{S}}

\algnewcommand{\Inputs}[1]{%
  \State \textbf{inputs:}\hspace*{\algorithmicindent}\parbox[t]{.8\linewidth}{\raggedright #1}
}
\algnewcommand{\Outputs}[1]{%
   \State \textbf{outputs:}\hspace*{\algorithmicindent}\parbox[t]{.8\linewidth}{\raggedright #1}
}
\algnewcommand{\Initialize}[1]{%
  \State \textbf{initialize:}
  \Statex \hspace*{\algorithmicindent}\parbox[t]{.8\linewidth}{\raggedright #1}
}
\makeatletter
\newcommand{\thickhline}{%
    \noalign {\ifnum 0=`}\fi \hrule height 1pt
    \futurelet \reserved@a \@xhline
}
\newcolumntype{"}{@{\hskip\tabcolsep\vrule width 1pt\hskip\tabcolsep}}
\makeatother
\newcommand{\mr}{\mathrm}
\newcommand{\BE}{\begin{equation}}
\newcommand{\EE}{\end{equation}}
\newcommand{\BS}{\begin{subequations}}
\newcommand{\ES}{\end{subequations}}

\begin{document}
\topmargin=0mm
\title{D-OAMP: A Denoising-based Signal Recovery Algorithm\\ for Compressed Sensing}
\name{Zhipeng Xue$^{\star}$ \qquad Junjie Ma$^{\dagger}$ \qquad Xiaojun Yuan$^{\star}$}
\address{$^{\star}$ School of Information Science and Technology, ShanghaiTech University \\
    $^{\dagger}$Department of Electronic Engineering, City University of Hong Kong}

\maketitle

\begin{abstract}
Approximate message passing (AMP) is an efficient iterative signal recovery algorithm for compressed sensing (CS). For sensing matrices with independent and identically distributed (i.i.d.) Gaussian entries, the behavior of AMP can be asymptotically described by a scaler recursion called state evolution. Orthogonal AMP (OAMP) is a variant of AMP that imposes a divergence-free constraint on the denoiser. In this paper, we extend OAMP to incorporate generic denoisers, hence the name D-OAMP. Our numerical results show that state evolution predicts the performance of D-OAMP well for generic denoisers when i.i.d. Gaussian or partial orthogonal sensing matrices are involved. We compare the performances of denosing-AMP (D-AMP) and D-OAMP for recovering natural images from CS measurements. Simulation results show that D-OAMP outperforms D-AMP in both convergence speed and recovery accuracy for partial orthogonal sensing matrices.
\end{abstract}

\endkeywords
%==============================================================
\section{Introduction}

Compressed sensing (CS) is a new paradigm for signal acquisition \cite{Donoho2006}. The CS measurement process can be modeled as
\BE
\y=\A\x_0+\w,
\EE
where $\A\in\mathbb{R}^{M\times N}$ is the sensing (or measurement) matrix, $\x_0$ the signal and $\w\sim\mathcal{N}(\bm{0},\sigma^2\bm{I})$ the measurement noise. This paper is concerned with the signal recovery problem, namely, estimating $\x_0$ from $\y$.

Approximate message passing (AMP) \cite{Donoho2009} is a low-cost iterative signal recovery algorithm. There are two modules involved in AMP: a linear filter and a component-wise nonlinear processor. The linear filter estimates the signal from compressed measurements, and the nonlinear processor refines the estimate by exploiting structural information of the signal (e.g., sparsity). A distinctive feature about AMP is that, when $\bm{A}$ has i.i.d. entries, the linear filter output can be modeled as an observation of the signal corrupted by white Gaussian noise \cite{Donoho2009,Bayati2011}. Furthermore, the variance of the Gaussian noise can be tracked through a scalar recursion called state evolution (SE) \cite{Donoho2009,Bayati2011}. The SE framework is useful for algorithm tuning. For instance,  \cite{guo2015near} develops a parametric AMP algorithm where the nonlinear processor is a linear combination of multiple elementary nonlinear functions, see also \cite{blu2007sure}. Based on SE analysis, the combination coefficients can be adaptively optimized in each iteration \cite{guo2015near}.

The original AMP algorithm mainly focuses on sparse signal recovery. To exploit more general signal structure, \cite{metzler2014denoising} proposed a denoising-based AMP (D-AMP) framework where generic denoisers are used as the nonlinear processor. A large number of popular image denoisers were examined in \cite{metzler2014denoising}. The numerical results showed that D-AMP with off-the-shelf denoisers achieve state-of-the-art recovery performance for i.i.d. Gaussian sensing matrices. Due to complexity considerations, it is also of interest to use sensing matrices constructed from fast implementable operators like discrete cosine transform (DCT). A potential problem for D-AMP with such partial orthogonal matrix is that the SE framework might be inaccurate and the recovery performance cannot be guaranteed. In \cite{perelli2015compressive},  the authors adopted the turbo recovery algorithm in \cite{ma2014turbo} to incorporate image denoisers such as BM3D \cite{dabov2006image}. However, the results in \cite{perelli2015compressive} showed that their proposed BM3D-turbo is inferior to D-AMP in reconstruction accuracy.

%\section{Related Works}
Orthogonal AMP (OAMP) \cite{ma2016oamp} is an extension of the turbo signal recovery algorithm in \cite{ma2014turbo}, and can be viewed as a special case of AMP where the denoiser is divergence-free. It was shown in \cite{ma2014turbo,ma2016oamp} that OAMP outperforms AMP for partial orthogonal sensing matrices when the entries of $\x_0$ are independently drawn from a known distribution. However, this random signal model does not hold for image processing applications considered in this paper.

In this paper, we propose a denoising-based OAMP algorithm (D-OAMP) for recovering natural images from CS measurements. D-OAMP generalizes OAMP by integrating generic denoisers. We consider both i.i.d. Gaussian sensing matrices and partial orthogonal matrices. We develop an SE framework for D-OAMP. Our numerical results show that the SE for D-OAMP is accurate for both i.i.d. Gaussian and partial orthogonal sensing matrices. Unlike \cite{perelli2015compressive}, our proposed D-OAMP outperforms D-AMP in both convergence speed and recovery accuracy when a partial orthogonal sensing matrix is involved and the BM3D denoiser is adopted.

\section{Denoising-based AMP}
\subsection{D-AMP Algorithm}
%The AMP Algorithm is a fast iterative algorithm proposed by Donoho, Arian Maleki, and Andrea Montanari\cite{donoho2010message}.

\begin{algorithm}
\caption{D-AMP algorithm}\label{algorithm1}
\begin{algorithmic}[1]
\Inputs{$\A^{M\times N}, \y^{M\times 1}, \hat \x_0=0, \z_0=\y$}
%\Initialize{$\hat \x_0=0, \z_0=\y$}
\For {$t=0, 1, 2, 3,\dots, k $}
\State $\hat\sigma_{t}^2=\frac{\|\z_{t}\|^2}{M}$
\State $\r_t=\x_t+\A^T \z_t$
\State $\hat \x_{t+1}=D_t(\r_{t}, \hat\sigma_t)$
\State $\z_{t+1}=\y-\A \hat \x_{t+1}+\z_t  \mr{div}\{D_t(\r_t)\}/M$
\EndFor
\Outputs{$\hat \x_{out}=\hat \x_k$}
\end{algorithmic}
\end{algorithm}

 Before presenting D-OAMP, we first review the D-AMP algorithm in \cite{metzler2014denoising}. The details of D-AMP are described in Algorithm \ref{algorithm1}.

 In D-AMP, $D(\r_t)$ represents a denoiser.\footnote{In general, the noise variance $\hat{\sigma}_t^2$ is a parameter of the denoiser. However, we sometimes write $D(\r_t)$ instead of $D_t(\r_{t}, \hat\sigma_t)$ for notational brevity.} From \cite{metzler2014denoising} the divergence of $D(\r_t)$ is defined as
 \BE \label{Eqn:divergence}
 \mr{div}\left\{D(\r_t)\right\}\equiv\frac{1}{N}\sum_{i=1}^N\frac{\partial \left(D(\r_t)\right)_i}{\partial r_i^t}. 
 \EE
 
%The last item in Step 6 of Algorithm \ref{algorithm1} is the so-called Onsager term \cite{Donoho2009}, and it plays a key role in D-AMP.
Many denoisers do not have explicit expressions. In such cases, we can use the following Monte Carlo method \cite{metzler2014denoising} to compute the divergence. Let $\bm{e} \sim \mathcal{N}(\bm{0},\bm{I})$ be an i.i.d. random Gaussian vector. The divergence of $D(\r^t)$ can be estimated as \cite{metzler2014denoising}
 \begin{align} \label{Eqn:divergence_app}
 	\begin{split}
 	 \mr{div}\left\{D(\r^t)\right\}\approx E_{\bm{e}}\left\{\bm{e}^T\left(\frac{D(\r^t+\delta \bm{e})-D(\r^t)}{\delta}\right)\right\},
	\end{split}
\end{align}
where $\delta$ is a small constant. The expectation in \eqref{Eqn:divergence_app} can be approximated by sample average. It is observed in \cite{metzler2014denoising} that one sample is good enough for high-dimensional problems.

\subsection{State Evolution}\label{Sec:SE-D-AMP}

When $\bm{A}$ has i.i.d. Gaussian entries, it is found empirically \cite{metzler2014denoising} that the mean square error (MSE) of D-AMP in each iteration can be characterized by the following state evolution (SE) recursion
\BS
\begin{align}
\tau_t^2 &= \frac{1}{\delta} v_t^2+\sigma^2,\\
v_{t+1}^2&=\frac{1}{N}E\left\{ \left\| D_t\left(\x_0+\tau_t\bm{e}\right)-\x_0\right\|^2\right\},
\end{align}
\ES
where the expectation is taken over $\bm{e}\sim\mathcal{N}(\bm{0},\bm{I})$. In the first iteration, $v_0=\|\bm{x}_0\|^2/N$.

The intuition behind the state evolutions is that $\bm{r}^t$ is an effective observation for $\bm{x}_0$ with additive white Gaussian noise (AWGN) \cite{Donoho2009}. This is rigorously proved in \cite{Bayati2011} for AMP when $\bm{A}$ is an i.i.d. Gaussian sensing matrix. For D-AMP, the validity of this property is examined through extensive simulations \cite{metzler2014denoising}. This ``effective AWGN observation'' property is attributed to the Onsager term \cite{Donoho2009,metzler2014denoising} $\z_t  \mr{div}\{D_t(\r_t)\}/m$ (See line 7 in Algorithm \ref{algorithm1}).

\section{Denoising-based OAMP}
We now present the proposed denoising-based OAMP (D-OAMP) algorithm.

\begin{algorithm}
\caption{D-OAMP algorithm}\label{algorithm2}
\begin{algorithmic}[1]
\Inputs{$\A^{M\times N}, \y^{M\times 1}, \sigma^2, \hat \x_0=0, \z_0=\y$}
%\Initialize{$\hat \x_0=0, \z_0=\y$}
\For {$t=0, 1, 2, 3\dots, k$}
\State $\hat v_{t}=\frac{\|\z_t\|^2-M \sigma^2}{tr(\A^T \A)}$
\State $\r_{t}=\hat \x_t+ \W_t \z_t$
\State $\hat \sigma_t^2=\Phi(\hat v_t^2)$
\State $\hat \x_{t+1}=D_t(\r_{t}, \hat \sigma_t)$
\State $\z_{t+1}=\y-\A \hat \x_{t+1}$
\EndFor
\Outputs{$\hat \x_{out}=D_{k}^{out}(\hat \r_{k}; \theta_{k})$}
\end{algorithmic}
\end{algorithm}

\subsection{D-OAMP Algorithm}
The D-OAMP algorithm is shown in Algorithm \ref{algorithm2}. In Algorithm \ref{algorithm2}, $\W_t$  is a linear filter. Readers are referred to \cite[Section III-A]{ma2016oamp} for details. $\Phi(\cdot)$ (which depends on $\bm{A}$ and $\bm{W}_t$) \cite[Eqn.~(31)]{ma2016oamp} is used to estimate the noise variance of $\r_t$. Different from D-AMP, the denoiser $D(\bm{r}_t)$ is required to be divergence-free \cite{ma2016oamp}, i.e. $\mr{div}\{D(\bm{r}_t)\}=0$.
(We will discuss how to construct divgence-free denoisers in Section \ref{Sec:divergence-free}.) The estimate of D-OAMP is produced by ${D}^{out}_t(\bm{r}_t)$, which is not restricted to be divergence-free \cite{ma2016oamp}.
%\ref{Sec:divergence-free}.
%\BE
%\mr{div}\{D(\bm{r}_t)\}=0.
%\EE
%The MSE estimators. Choices of linear filters (omit detailed expression)
\subsection{State Evolution}\label{Sec:SE-D-OAMP}
Following \cite{ma2016oamp}, we use the SE recursion below to characterize D-OAMP
\BS
\begin{align}\label{Equ:SE}
\tau_t^2 &=\Phi(v_t^2),\\
v_{t+1}^2&=\frac{1}{N}E\left\{ \left\| D_t\left(\bm{x}_0+\tau_t\bm{e}\right)-\bm{x}_0\right\|^2\right\},
\end{align}
\ES
where the expectation is taken over $\bm{e}\sim\mathcal{N}(\bm{0},\bm{I})$, and $v_0=\|\bm{x}_0\|^2/N$. Notice that the final estimate of D-OAMP is produced by ${D}_t^{out}$ (which is not restricted to be divergence-free), and the corresponding MSE is predicted as
\BE
\frac{1}{N}E\left\{ \left\| D_t^{out}\left(\bm{x}_0+\tau_t\bm{e}\right)-\bm{x}_0\right\|^2\right\}.
\EE
We will show that SE for D-OAMP is accurate for both i.i.d. Gaussian sensing matrices and partial orthogonal sensing matrices (with appropriate randomization). In comparison, the SE of D-AMP in Section \ref{Sec:numerical} can be inaccurate for partial orthogonal matrices.

\subsection{Divergence-free Denoisers for OAMP}\label{Sec:divergence-free}
%We now discuss the details about constructing a divergence-free denoiser.
\subsubsection{Construction of divergence-free denoisers}
The state evolutions in Section \ref{Sec:SE-D-AMP} and Section \ref{Sec:SE-D-OAMP} are based on the AWGN model $\bm{r}=\bm{x}_0+\tau\bm{e} \label{Eqn:AWGN}$,
%\BE \label{Eqn:AWGN}
%\bm{r}=\bm{x}_0+\tau\bm{e},
%\EE
where $\bm{e}\sim\mathcal{N}(\bm{0},\bm{I})$. For brevity, we omit the iteration index $t$ for $\bm{r}$.
Given a denoiser $\hat{D}(\bm{r})$, we construct a divergence-free denoiser as follows \cite{ma2016oamp}
\BE \label{Eqn:divergence-free}
D(\bm{r})=C\left( \hat{D}(\r)-\mr{div}\left\{\hat{D}(\r)\right\} \r \right),
\EE
where $C$ is an arbitrary constant. From \eqref{Eqn:divergence}, the divergence of $D(\bm{r})$ in \eqref{Eqn:divergence-free} can be expressed as
\begin{align} \label{Eqn:divergence-free-2b}
\begin{split}
&\mr{div}\{D(\bm{r})\}=\\
&C\left( \mr{div}\{\hat{D}(\bm{r})\}  -\mr{div}\{\hat{D}(\bm{r})\}  -\frac{1}{N}\sum_{i=1}^N\frac{\partial \mr{div}\{\hat{D}(\bm{r})\}}{\partial r_i} r_i  \right)\\
&\approx 0,
\end{split}
\end{align}
where we assume that the contribution of each individual $r_i$ in $\mr{div}\{\hat{D}(\bm{r})\}$ is negligible for a high-dimensional problem, and so the third term in \eqref{Eqn:divergence-free-2b} is approximately zero. 

To see the rationale behind \eqref{Eqn:divergence-free-2b}, consider the following special case. Let $\bm{x}_0$ be an i.i.d. vector and $\hat{D}(\cdot)$ a component-wise function of $\bm{r}$ (i.e., $(\hat{D}(\bm{r}))_i=\hat{D}(r_i)$). In this case, $\mr{div}\{\hat{D}(\bm{r})\}$ converges to a constant (independent of each individual $r_i$) as $N\to\infty$ by the law of large numbers. For non-random $\bm{x}_0$ and generic $\hat{D}$, we still expect $\mr{div}\{\hat{D}(\bm{r})\}$ to be self-averaging, although this may not be true in a strict sense. Nevertheless, our numerical results in Section \ref{Sec:numerical} show that the SE framework developed based on this assumption is reasonably accurate for real images and practical denoisers.

%\par\vspace{3pt}
%\textit{2) Parameter tuning based on SURE}\vspace{3pt}
\subsubsection{Parameter tuning based on SURE}
For the divergence-free denoiser in \eqref{Eqn:divergence-free}, $C$ is a free-parameter. Ideally, we choose $C$ that minimizes
\BE \label{Eqn:MSE}
\mr{MSE}=\frac{1}{N}{E}\left\{\| D(\bm{r})-\bm{x}_0\|^2 \right\},
\EE
where $\bm{r}=\bm{x}_0+\tau\bm{e}$ and the expectation in \eqref{Eqn:MSE} is with respect to $\bm{e}$. The problem is that we cannot compute the MSE in \eqref{Eqn:MSE} since $\bm{x}_0$ is unknown. Following \cite{guo2015near,blu2007sure}, we choose $C$ that minimizes Stein's unbiased risk estimate (SURE) \cite{Stein1981}
\BS \label{Eqn:SURE}
\begin{align}
\widehat{\mr{MSE}}&=\frac{1}{N}\|D(\r)-\r\|^2+\frac{2\tau^2}{N}\mr{div}\{D(\r)\}-\tau^2\\
&=\frac{1}{N}\|D(\r)-\r\|^2-\tau^2\\
&=\frac{1}{N}\|C\left( \hat{D}(\r)-\mr{div}\{\hat{D}(\r)\} \r \right)-\r\|^2-\tau^2
\end{align}
\ES
where the second equality is due to the fact that $D(\r)$ is divergence-free (c.f., \eqref{Eqn:divergence-free-2b}). The optimal $C$ that minimizes \eqref{Eqn:SURE} (which is a quadratic function of $C$) is given by
\BE \label{Eqn:C_opt}
C^{\star}=\frac{\r^T\left(   \hat{D}(\r)-\mr{div}\{\hat{D}(\r)\} \r\right)}{\| \hat{D}(\bm{r})-\mr{div}\{\hat{D}(\bm{r})\} \r\|^2}.
\EE
%\par\vspace{3pt}
%\textit{3) The SURE-LET framework}\vspace{3pt}\par
\subsubsection{The SURE-LET framework}
The SURE-LET framework in \cite{guo2015near,blu2007sure} can also be integrated into the D-OAMP algorithm. In SURE-LET, the denoiser $D(\r)$ is given by a linear combination of multiple elementary denoisers:
\BE
D(\r)=\sum_{k=1}^KC_k\left(\hat{D}_k(\r)-\mr{div}\{\hat{D}_k(\r)\} \r\right).
\EE
Here, each elementary denoiser is divergence-free (c.f., \eqref{Eqn:divergence-free}), and so $D(\r)$ is also divergence-free. \par
Denote $\bm{C}^{\star}\equiv[C_1^{\star},C_2^{\star},\ldots,C_K^{\star}]^T$ where $\{C_k^{\star}\}$ are the optimal values of $\{C_k\}$ that minimize the SURE. Let $\bm{G}_k\equiv\hat{D}_k(\r)-\mr{div}\{\hat{D}_k(\r)\} \r$, and define $M_{i,j}\equiv \bm{G}_i^T\bm{G}_j$, $\bm{b}\equiv [\bm{G}_1^T\r,\bm{G}_2^T\r,\ldots,\bm{G}_K^T\r]^T$. Following similar derivations as \eqref{Eqn:C_opt}, we obtain the following optimal combining coefficients:
\begin{align}
\bm{C}^{\star}=\bm{M}^{-1}\bm{b}.
\end{align}

%\subsection{State Evolution}\label{Sec:SE-D-OAMP}
%Following \cite{ma2016oamp}, we use the following SE recursion to characterize D-OAMP
%\BS
%\begin{align}\label{Equ:SE}
%\tau_t^2 &=\Phi(v_t^2),\\
%v_{t+1}^2&=\frac{1}{N}E\left\{ \left\| D\left(\bm{x}_0+\tau_t\bm{e}\right)-\bm{x}_0\right\|^2\right\},
%\end{align}
%\ES
%where the expectation is taken over $\bm{e}\sim\mathcal{N}(\bm{0},\bm{I})$, and $v_0=\|\bm{x}_0\|^2/N$. Notice that the final estimate of D-OAMP is produced by ${D}_t^{out}$ (which is not restricted to be divergence-free), and the corresponding MSE is predicted as
%\BE
%\frac{1}{N}E\left\{ \left\| D_t^{out}\left(\bm{x}_0+\tau_t\bm{e}\right)-\bm{x}_0\right\|^2\right\}.
%\EE
%We will show that SE for D-OAMP is accurate for both i.i.d. Gaussian sensing matrices and partial orthogonal sensing matrices (with appropriate randomization). In comparison, the SE of D-AMP in Section \ref{Sec:numerical} can be inaccurate for partial orthogonal matrices.

%-----------------------------beginfigure-------------------------
\begin{figure*}[!t]
    \centering
    \begin{subfigure}[b]{0.3\textwidth}
    \centering
        \includegraphics[trim=26 190 70 205,clip,width=\textwidth]{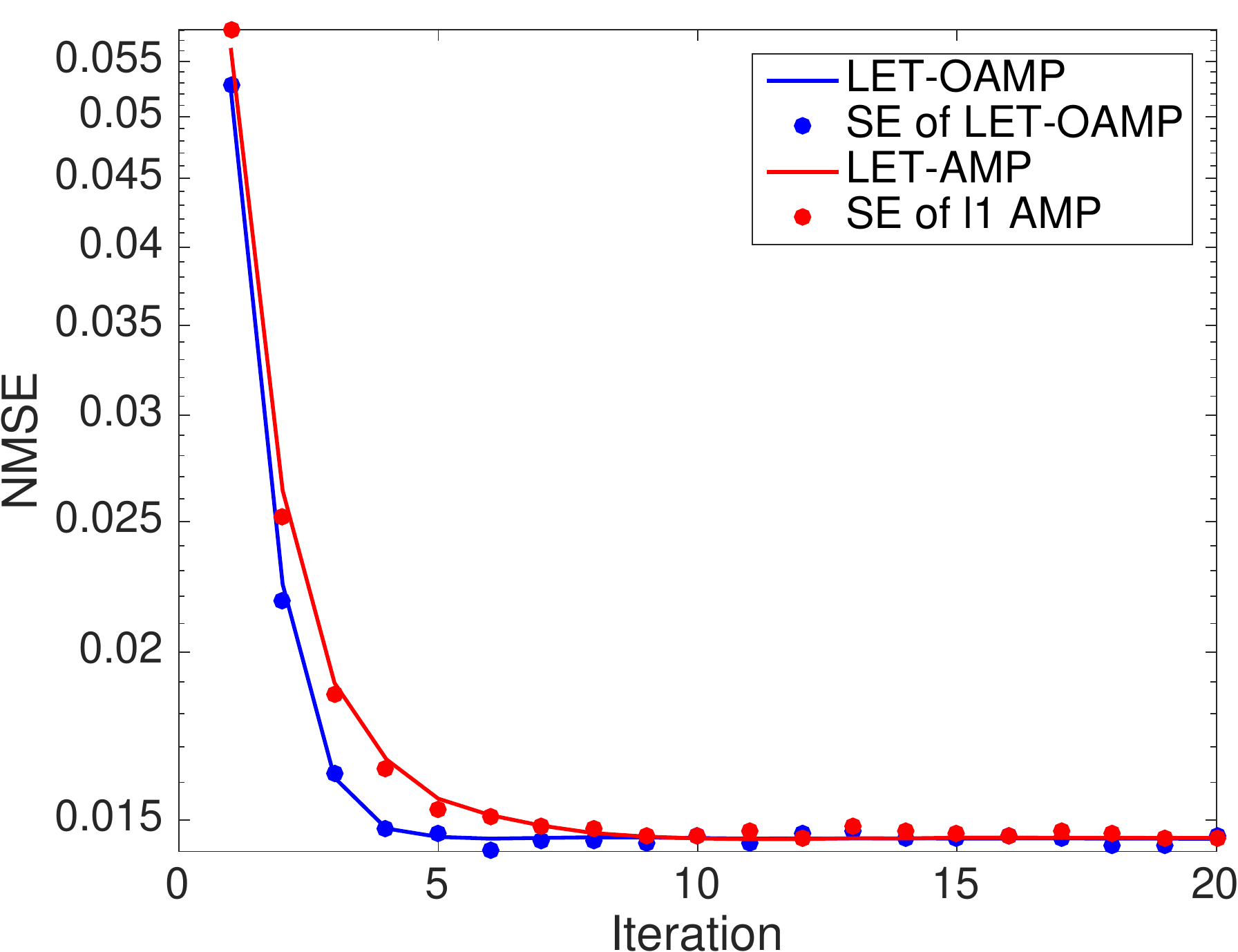}
        \caption{Gaussian matrix, measurement rate 0.3, input noiseless image Fingerprint.}
        \label{fig:a}
    \end{subfigure}
    ~
    \begin{subfigure}[b]{0.3\textwidth}
    \centering
    \includegraphics[trim=30 190 65 205,clip,width=\textwidth]{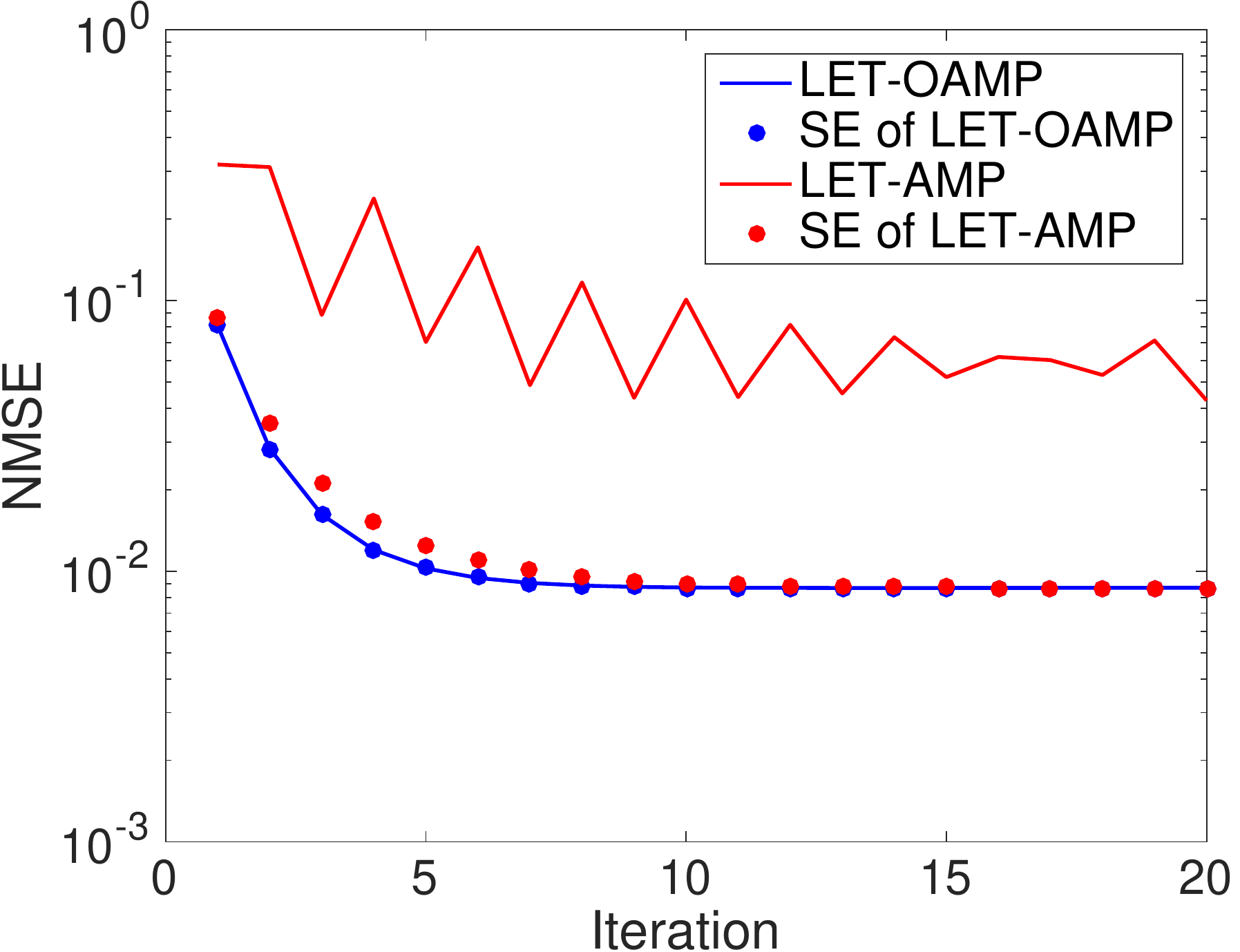}
        \caption{Orthogonal matrix, measurement rate 0.3, input noiseless image Fingerprint.}
        \label{fig:b}
    \end{subfigure}
    ~
    \begin{subfigure}[b]{0.3\textwidth}
    \centering
        \includegraphics[trim=20 190 65 205,clip,width=\textwidth]{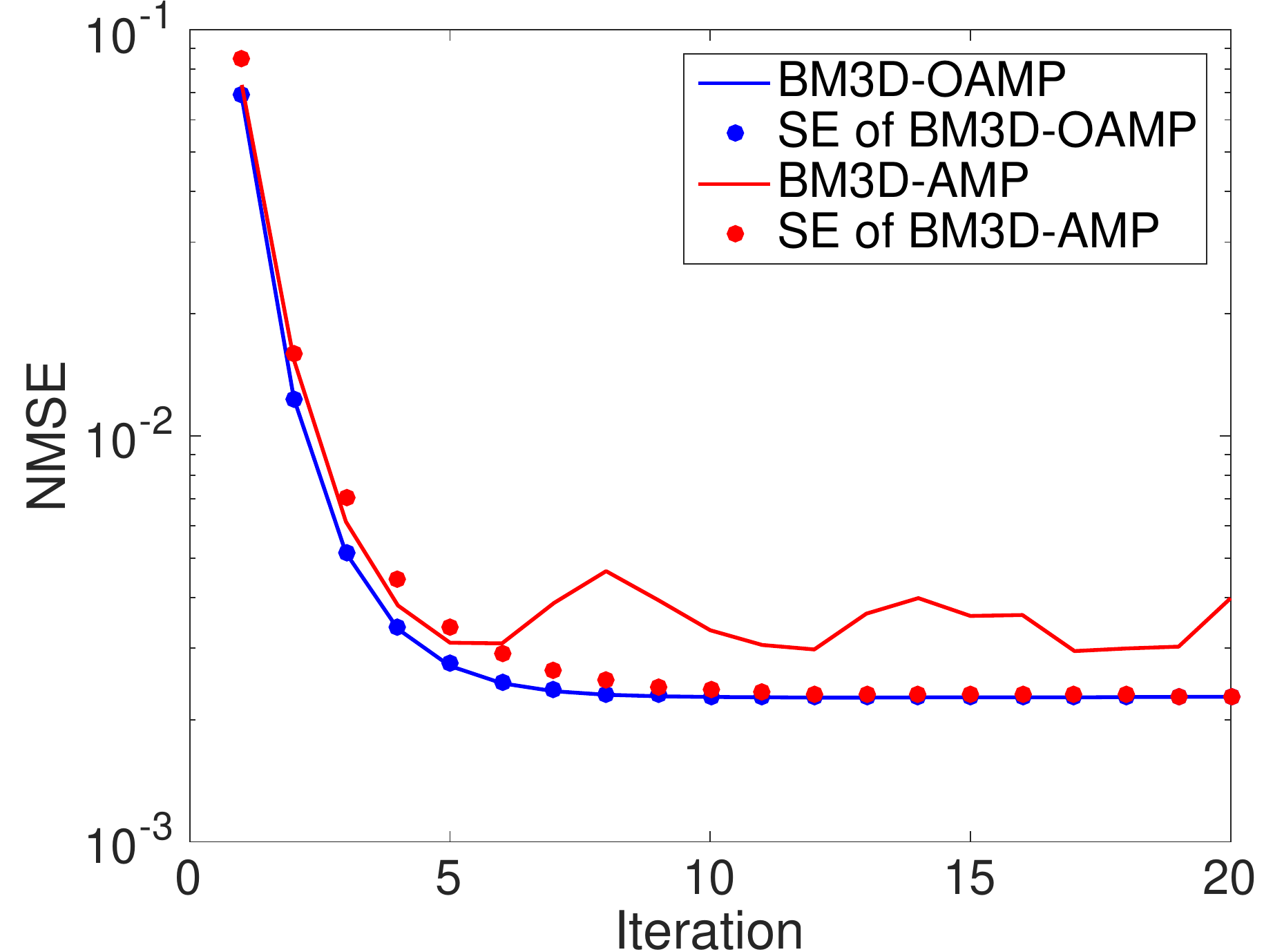}
        \caption{Orthogonal matrix, measurement rate 0.3, input noiseless image Fingerprint.}
        \label{fig:c}
    \end{subfigure}
     \caption{State evolution of D-AMP and D-OAMP}\label{SE}
\end{figure*}
%-------------------------tables---------------------------------
\begin{table*}[!ht]
\centering
\begin{tabular}{l|lll|lll|lll|lll}
\thickhline
\textbf{Images} & \multicolumn{3}{c|}{\textbf{Lena}}                                                                           & \multicolumn{3}{c|}{\textbf{Boat}}   & \multicolumn{3}{c|}{\textbf{Barbara}}                                                                        & \multicolumn{3}{c}{\textbf{Fingerprint}}                                                                   \\ \thickhline
    \textbf{Measurement rate}     & \multicolumn{1}{c}{\textbf{30\%}} & \multicolumn{1}{c}{\textbf{50\%}} & \multicolumn{1}{c|}{\textbf{70\%}} & \multicolumn{1}{c}{\textbf{30\%}} & \multicolumn{1}{c}{\textbf{50\%}} & \multicolumn{1}{c|}{\textbf{70\%}} & \multicolumn{1}{c}{\textbf{30\%}} & \multicolumn{1}{c}{\textbf{50\%}} & \multicolumn{1}{c|}{\textbf{70\%}} & \multicolumn{1}{c}{\textbf{30\%}} & \multicolumn{1}{c}{\textbf{50\%}} & \multicolumn{1}{c}{\textbf{70\%}} \\ \thickhline
EM-GM-AMP  & 26.89                              & 29.50                              & 32.38                              & 24.70                              & 27.78                              & 30.95     &24.47                              & 27.69                              & 32.14                              & 22.51                              & 26.04                              & 29.58                         \\\hline
LET-AMP   &22.53           & 30.78                              & 33.74                              &21.51                & 29.00                              & 32.42       &19.69                         & 28.92                              & 32.46                              & 17.48                    & 27.25                              & 31.11                        \\
LET-OAMP    & \textbf{27.72}                              & \textbf{31.12}                              & \textbf{34.80}                              & \textbf{25.70}                              & \textbf{29.44}                              & \textbf{33.54}  & \textbf{25.49}                              & \textbf{29.36}                              & \textbf{33.70}                              & \textbf{23.60}                              & \textbf{27.82}                              & \textbf{32.48} \\\hline
  BM3D-AMP   & 34.72                                  & 38.20                              & 38.98                              & 34.09                                  & 34.70                              & 37.92       & 35.15                                 & 38.24                              & 41.39                              & 28.97                                 & 33.18                              & 37.00                          \\
BM3D-OAMP     & \textbf{35.50}                              & \textbf{38.26}                              & \textbf{42.36}                              & \textbf{34.59}                              & \textbf{37.60}                              & \textbf{41.28}             & \textbf{36.51}                              & \textbf{39.83}                              & \textbf{43.35}                              & \textbf{31.01}                              & \textbf{35.24}                              & \textbf{39.79}                 \\ \thickhline
\end{tabular}
\caption{PSNR of reconstructed images under orthogonal matrix}
\label{table1}
\end{table*}

\begin{table*}[!ht]
\centering
\begin{tabular}{l|lll|lll|lll|lll}
\thickhline
\textbf{Images} & \multicolumn{3}{c|}{\textbf{Lena}}                                                                           & \multicolumn{3}{c|}{\textbf{Boat}}   & \multicolumn{3}{c|}{\textbf{Barbara}}                                                                        & \multicolumn{3}{c}{\textbf{Fingerprint}}                                                                          \\ \thickhline
\textbf{Measurement rate}    & \multicolumn{1}{c}{\textbf{30\%}} & \multicolumn{1}{c}{\textbf{50\%}} & \multicolumn{1}{c|}{\textbf{70\%}} & \multicolumn{1}{c}{\textbf{30\%}} & \multicolumn{1}{c}{\textbf{50\%}} & \multicolumn{1}{c|}{\textbf{70\%}}& \multicolumn{1}{c}{\textbf{30\%}} & \multicolumn{1}{c}{\textbf{50\%}} & \multicolumn{1}{c|}{\textbf{70\%}} & \multicolumn{1}{c}{\textbf{30\%}} & \multicolumn{1}{c}{\textbf{50\%}} & \multicolumn{1}{c}{\textbf{70\%}}  \\ \thickhline
LET-AMP   &12.68           & 12.47                              & 12.29                              &12.48                & 12.7                              &  12.75        &12.49                         & 12.45                              & 12.83                              & 12.74                    & 12.41                              & 13.57                  \\
LET-OAMP    & \textbf{4.60}                              & \textbf{3.97}                              & \textbf{2.72}                              & \textbf{4.39}                              & \textbf{3.86}                              & \textbf{3.33} & \textbf{4.32}                              & \textbf{3.72}                              & \textbf{3.45}                              & \textbf{7.95}                              & \textbf{5.1}                              & \textbf{4.92} \\
 \thickhline
\end{tabular}
\caption{Reconstruction time of images under orthogonal matrix}
\label{table2}
\end{table*}
%-------------------------tables--------------------------------
%-----------------------------endfigure-------------------------
\section{Numerical Results}\label{Sec:numerical}

In this section, we provide numerical results of D-AMP and D-OAMP in CS image recovery. The recovery accuracy is measured by both normalized mean square error (NMSE) and peak signal-to-noise ratio (PSNR):
\begin{align}
	%\text{MSE}&=10\log_{10}\frac{\|\x_0\|^2}{\|\hat \x-\x_0\|^2}	\\
	\text{NMSE}&=\frac{\|\hat \x-\x_0\|^2}{\|\x_0\|^2},	\\
	\text{PNSR}&=10\log_{10}\left(\frac{\text{MAX}^2}{\text{MSE}}\right),
\end{align}
where MAX denotes the maximum pixel values of the image. We consider two denoisers, the piecewise linear kernel in \cite{guo2015near} and BM3D denoiser \cite{dabov2006image} to construct denoiser $D(\r)$ respectively. The corresponding algorithms are denoted respectively as LET-AMP, LET-OAMP, BM3D-AMP and BM3D-OAMP.
\subsection{Accuracy of State Evolution}
%We first examine the accuracy of SE prediction of D-OAMP.
\subsubsection{Gaussian sensing matrix}
Consider an i.i.d. Gaussian $\bm{A}$ with each $A_{i,j}\sim\mathcal{N}(0,1/M)$. In this case, we choose $\W_t$ to be the linear minimum mean square error estimation (LMMSE) matrix \cite[Eqn.~(38)]{ma2016oamp}. The $\Phi$ function in \eqref{Equ:SE} is given by \cite[Eqn.~(43c)]{ma2016oamp}
\begin{align}
	\begin{split}
		\Phi(v_t^2)= \frac{\sigma^2+c v_t^2+\sqrt{(\sigma^2+c v_t^2)^2+4\sigma^2v_t^2}}{2},
	\end{split}
\end{align}
where $c=(N-M)/M$. The simulated MSEs and SE predictions are shown in Fig.~\ref{fig:a}. We see that the SE predictions match well with simulated MSEs for both D-AMP and D-OAMP.
%---------------------
\subsubsection{Partial orthogonal sensing matrix}
We use the following partial orthogonal matrix:
\BE \label{Eqn:doubleDCT}
%\bm{A}=\sqrt{\frac{N}{M}} \S \F^T \bm{\Theta_1} \F\bm{\Theta_2},
\bm{A}=\sqrt{N/M} \S \F^T \bm{\Theta_1} \F\bm{\Theta_2},
\EE
where $\S$ is a selective matrix, $\F$ is a DCT matrix and $\bm{\Theta_1}, \bm{\Theta_2}$ are diagonal matrices with diagonals being 1 or -1 with equal probability. Here, $ \bm{\Theta_1}$ and $\bm{\Theta_2}$ are introduced to properly randomize $\bm{A}$. The SE in \eqref{Equ:SE} now becomes \cite[Eqn.~(45)]{ma2016oamp}
\begin{align}
	\begin{split}
		\Phi(v_t^2)= \frac{N-M}{M} v_t^2+\sigma^2.
	\end{split}
\end{align}
\par
Figs.~\ref{fig:b} and \ref{fig:c} plot, respectively, the MSE performances of LET-AMP/LET-OAMP and BM3D-AMP/BM3D-OAMP. In these simulations, we use the ``fingerprint" from the Javier Portilla’s dataset \cite{CIVtest}. More comprehensive numerical comparisons will be provided in Table \ref{table1}. We find that LET-AMP in Algorithm \ref{algorithm1} does not work well (diverge under the current scenario) for partial orthogonal matrices. To address this, we adopt more robust median estimator for $\hat\sigma^2$ proposed in \cite{anitori2013design} in D-AMP. Even after tuning, the D-AMP algorithm in both Figs.~\ref{fig:b} and \ref{fig:c} deviate from SE prediction and does not work well under low sampling rate (like Fig.~\ref{fig:b}). In contrast, the SE for D-OAMP well predicts the simulated MSEs.
\subsection{Performance Comparison}
In Table \ref{table1}, we report the PSNR of noiseless image recovery (of size $512\times 512$) under the orthogonal matrix given in \eqref{Eqn:doubleDCT}. We see that D-OAMP uniformly outperforms D-AMP and EM-GM-AMP \cite{vila2013expectation}. We further compare the reconstruction time of LET-AMP and LET-OAMP in Table \ref{table2}. Both algorithms are run until convergence is reached. We see that LET-OAMP also converges faster than LET-AMP. This is consistent with the observation in Fig.~\ref{fig:b}.
%The performance of D-OAMP under other partial orthogonal sensing matrixes like DCT and DFT matrix is also better than D-AMP, the results are similar and not provided here.

\section{Conclusion}
In this paper, we proposed D-OAMP algorithm for compressed sensing. D-OAMP does not require the knowledge of distribution of input signal and therefore can be adopted in many applications. Numerical results show that D-OAMP outperforms D-AMP in terms of both recovery accuracy and convergence speed for partial orthogonal sensing matrices.
%D-OAMP doesn't need to know the prior distribution of input signal, which can be adopt to many types of signals.
%==============================================================
\bibliographystyle{IEEEtran}	% (uses file "plain.bst")
%\bibliography{IEEEabrv,ref}		% Junjie

\begin{thebibliography}{10}
\providecommand{\url}[1]{#1}
\csname url@samestyle\endcsname
\providecommand{\newblock}{\relax}
\providecommand{\bibinfo}[2]{#2}
\providecommand{\BIBentrySTDinterwordspacing}{\spaceskip=0pt\relax}
\providecommand{\BIBentryALTinterwordstretchfactor}{4}
\providecommand{\BIBentryALTinterwordspacing}{\spaceskip=\fontdimen2\font plus
\BIBentryALTinterwordstretchfactor\fontdimen3\font minus
  \fontdimen4\font\relax}
\providecommand{\BIBforeignlanguage}[2]{{%
\expandafter\ifx\csname l@#1\endcsname\relax
\typeout{** WARNING: IEEEtran.bst: No hyphenation pattern has been}%
\typeout{** loaded for the language `#1'. Using the pattern for}%
\typeout{** the default language instead.}%
\else
\language=\csname l@#1\endcsname
\fi
#2}}
\providecommand{\BIBdecl}{\relax}
\BIBdecl

\bibitem{Donoho2006}
D.~L. Donoho, ``Compressed sensing,'' \emph{{IEEE} Trans. Inf. Theory},
  vol.~52, no.~4, pp. 1289--1306, Apr. 2006.

\bibitem{Donoho2009}
D.~L. Donoho, A.~Maleki, and A.~Montanari, ``Message-passing algorithms for
  compressed sensing,'' in \emph{Proc. Nat. Acad. Sci.}, vol. 106, no.~45, Nov.
  2009.

\bibitem{Bayati2011}
M.~Bayati and A.~Montanari, ``The dynamics of message passing on dense graphs,
  with applications to compressed sensing,'' \emph{{IEEE} Trans. Inf. Theory},
  vol.~57, no.~2, pp. 764--785, Feb. 2011.

\bibitem{guo2015near}
C.~Guo and M.~E. Davies, ``Near optimal compressed sensing without priors:
  Parametric {SURE} approximate message passing,'' \emph{{IEEE} Trans. Signal
  Process.}, vol.~63, no.~8, pp. 2130--2141, Mar. 2015.

\bibitem{blu2007sure}
T.~Blu and F.~Luisier, ``The {SURE-LET} approach to image denoising,''
  \emph{{IEEE} Trans. Image Process.}, vol.~16, no.~11, pp. 2778--2786, Nov.
  2007.

\bibitem{metzler2014denoising}
C.~A. Metzler, A.~Maleki, and R.~G. Baraniuk, ``From denoising to compressed
  sensing,'' \emph{{IEEE} Trans. Inf. Theory}, vol.~62, no.~9, pp. 5117--5144,
  Sept. 2016.

\bibitem{perelli2015compressive}
A.~Perelli and M.~E. Davies, ``Compressive computed tomography image
  reconstruction with denoising message passing algorithms,'' in \emph{Signal
  Processing Conference (EUSIPCO), 2015 23rd European}.\hskip 1em plus 0.5em
  minus 0.4em\relax IEEE, Dec. 2015, pp. 2806--2810.

\bibitem{ma2014turbo}
J.~Ma, X.~Yuan, and L.~Ping, ``Turbo compressed sensing with partial {DFT}
  sensing matrix,'' \emph{{IEEE} Signal Process. Lett.}, vol.~22, no.~2, pp.
  158--161, Feb. 2015.

\bibitem{dabov2006image}
K.~Dabov, A.~Foi, V.~Katkovnik, and K.~Egiazarian, ``Image denoising with
  block-matching and {3D} filtering,'' in \emph{Electronic Imaging 2006}.\hskip
  1em plus 0.5em minus 0.4em\relax International Society for Optics and
  Photonics, 2006, pp. 606\,414--606\,414.

\bibitem{ma2016oamp}
J.~Ma and L.~Ping, ``Orthogonal {AMP},'' \emph{arXiv preprint
  arXiv:1602.06509}, 2016.

\bibitem{Stein1981}
C.~M. Stein, ``Estimation of the mean of a multivariate normal distribution,''
  \emph{The annals of Statistics}, pp. 1135--1151, 1981.

\bibitem{CIVtest}
``{CIV Test Images},''
  \url{http://www.io.csic.es/PagsPers/JPortilla/image-processing/bls-gsm/63-test-images},
  accessed 2016-05-19.

\bibitem{anitori2013design}
L.~Anitori, A.~Maleki, M.~Otten, R.~G. Baraniuk, and P.~Hoogeboom, ``Design and
  analysis of compressed sensing radar detectors,'' \emph{{IEEE} Trans. Signal
  Process.}, vol.~61, no.~4, pp. 813--827, Feb. 2013.

\bibitem{vila2013expectation}
J.~P. Vila and P.~Schniter, ``Expectation-maximization {Gaussian}-mixture
  approximate message passing,'' \emph{{IEEE} Trans. Signal Process.}, vol.~61,
  no.~19, pp. 4658--4672, Oct. 2013.

\end{thebibliography}

\end{document}